\documentclass[ppreprint]{sigplanconf}

\usepackage{algpseudocode}
\usepackage{amsmath}
\usepackage{amssymb}
\usepackage{amsfonts}
\usepackage{comment}
\usepackage{graphicx}
\usepackage{multirow}
\usepackage{url}
\usepackage{xspace}
\usepackage{tikz}
\usetikzlibrary{arrows}
\usetikzlibrary{positioning}

\begin{document}

\conferenceinfo{SOAP '13}{June 20, 2013, Seattle, Washington, USA.} 
\copyrightyear{2013} 
\copyrightdata{ISBN 978-1-4503-2201-0/13/06}

\title{Interprocedural Data Flow Analysis in Soot using Value Contexts}

\authorinfo{Rohan Padhye}
           {Indian Institute of Technology Bombay}
           {rohanpadhye@iitb.ac.in}
\authorinfo{Uday P. Khedker}
           {Indian Institute of Technology Bombay}
           {uday@cse.iitb.ac.in}

\maketitle

\begin{abstract}

An interprocedural analysis is precise if it is flow sensitive and fully context-sensitive even 
in the presence of recursion. Many methods of interprocedural analysis sacrifice precision for 
scalability while some are precise but limited to only a certain class of problems.

Soot currently supports interprocedural analysis of Java programs using
graph reachability. However, this approach is restricted to IFDS/IDE problems,
and is not suitable for
general data flow frameworks such as heap reference analysis and 
points-to analysis which have non-distributive flow functions.

We describe a general-purpose interprocedural analysis
framework for Soot using data flow values for context-sensitivity. This
framework is not restricted to problems with distributive flow
functions, although the lattice must be finite. It combines the
key ideas of the tabulation method of the functional approach and the technique of
value-based termination of call string construction. 

The efficiency and precision of interprocedural analyses is heavily 
affected by the precision of the underlying call graph. This is especially 
important for object-oriented languages like Java where
virtual method invocations cause an explosion of spurious call edges if
the call graph is constructed naively. We have instantiated our framework with a flow
and context-sensitive \text{points-to} analysis in Soot, which enables the construction
of call graphs that are far more precise than those constructed by Soot's 
\textsc{spark} engine.

\end{abstract}




\section{Introduction}

Interprocedural data flow analysis incorporates the effects of procedure
calls on the callers and callees. A context-insensitive analysis does not 
distinguish between distinct calls to a procedure. 
This causes the propagation of data flow values
across interprocedurally invalid paths (i.e. paths in which
calls and returns may not match) resulting in a loss of  precision. A
context-sensitive analysis restricts the propagation to valid paths and
hence is more precise.

Two most general methods of precise flow and context-sensitive analysis
are the {\em Functional\/} approach and the {\em Call Strings\/} approach~\cite{call-strings}.
The functional
approach constructs summary flow functions for procedures by reducing
compositions and meets of flow functions of individual statements 
to a single flow function, which is used directly in call statements. However,
constructing summary flow functions may not be possible in general. 
The tabulation method of the functional approach overcomes this restriction 
by enumerating the functions as pairs of input-output data flow values for each procedure,
but requires a finite lattice.

The call strings method remembers calling contexts in terms of unfinished calls 
as call strings. However, it requires 
an exponentially large number of call strings. 
The technique of value based termination of call string
construction~\cite{cs-vbt} uses data flow values to restrict the combinatorial explosion of contexts and
improves the efficiency significantly without any loss of precision. 

Graph reachability based interprocedural analysis~\cite{ifds,ths-ide} is
a special case of the functional approach. Formally, it requires flow
functions \text{$2^A\mapsto 2^A$} to distribute over the meet
operation so that they can be decomposed into meets of flow functions
\text{$A\mapsto A$}. Here $A$ can be either a finite set $D$ (for IFDS
problems~\cite{ifds}) or a mapping \text{$D\mapsto L$} (for IDE problems~\cite{ths-ide}) from a
finite set $D$ to a lattice of values $L$. Intuitively, $A$ represents
a node in the graph and a function \text{$A\mapsto A$} decides the
nature of the edge from the node representing the argument to the node
representing the result. Flow function composition then reduces to a
transitive closure of the edges resulting in paths in the graph.

The efficiency and precision of interprocedural analyses is heavily
affected by the precision of the underlying call graph. This is especially 
important for object oriented languages like Java where
virtual method invocations cause an explosion of spurious call edges if
the call graph is constructed naively.

Soot~\cite{soot} has been a stable and popular choice for hundreds of client 
analyses for Java programs, though it has traditionally lacked an interprocedural
framework. Bodden \cite{ifds-soot} has recently implemented support for 
interprocedural analysis using graph reachability. 
The main limitation of this approach is that it is not suitable for
general data flow frameworks with non-distributive flow functions
such as heap reference analysis or points-to analysis.
For example, consider the Java statement
\texttt{x = y.n} to be processed for points-to analysis. If we have points-to edges
\text{$y \rightarrow o_1$} and \text{$o_1 . n \rightarrow o_2$} before the statement 
(where $o_1$ and $o_2$ are heap objects), then it is not possible to 
correctly deduce that the edge \text{$x \rightarrow o_2$}
should be generated after the statement if we consider each input edge independently.
The flow function for this statement is a function of the points-to graph as
a whole and cannot be decomposed into independent functions of each edge and then merged
to get a correct result.

We have implemented a generic framework for performing flow and context-sensitive 
interprocedural data flow analysis that does not require
flow functions to be distributive. However, the flow functions must be
monotonic and the lattice of data flow values must be finite. 
The framework uses value-based contexts and is an adaptation of the tabulation method of the
functional approach and the modified call strings
method. 

Our implementation is agnostic to any analysis toolkit or intermediate representation
as it is parameterized using generic types.
Since our core classes are similar to the intra-procedural
framework of Soot, it integrates with Soot's Jimple IR seamlessly.

We have instantiated our framework with a flow and context-sensitive points-to 
analysis in Soot, which enables the construction of call graphs that are far 
more precise than those constructed by Soot's \textsc{spark} engine.

The rest of the paper is organized as follows: Section~\ref{sec-background}
describes our method. Section~\ref{sec-framework} outlines the API of our
implementation framework. Section~\ref{sec-results} presents the results of 
call graph construction. Finally, Section~\ref{sec-conclusion} concludes the paper by
describing the current status and future possibilities of our work.

\section{Interprocedural Analysis Using Value Contexts}
\label{sec-background}  

The tabulation method of the functional approach~\cite{call-strings}
and the modified call strings approach~\cite{cs-vbt} both revolve
around the same key idea: if two or more calls to a
procedure $p$ have the same the data flow value (say $x$) at the entry
of $p$, then all of them will have an identical data flow value (say $y$)
at the exit of $p$. 
The tabulation method uses this idea to
enumerate flow functions in terms of pairs of input-output values
\text{$(x,y)$} whereas the modified call strings method uses it to
partition call strings based on input values, reducing
the number of call strings significantly. 

\newcommand{\method}{\text{\sf\em method}\xspace}
\newcommand{\enval}{\text{\sf\em entryValue}\xspace}
\newcommand{\exval}{\text{\sf\em exitValue}\xspace}

The two methods lead to an important conclusion: Using data flow values
as contexts of analysis can avoid re-analysis of procedure bodies. We
make this idea explicit by defining a \text{\textit{value context}} \text{$X
= \langle \method, \enval\rangle$}, where \enval is the data flow
value at the entry to a procedure \method. Additionally, we define
a mapping \text{$\exval(X)$} which gives the data flow value at the
exit of \method. As data flow analysis is an iterative process,
this mapping may change over time (although it will follow a descending chain
in the lattice). The new value is propagated
to all callers of \method if and when this mapping changes. With this
arrangement, intraprocedural analysis can be performed for each value
context independently, handling flow functions in the usual way; only
procedure calls need special treatment.

Although the number of value contexts created per procedure 
is theoretically proportional to the size of the lattice in the worst-case, 
we have found that in practice the number of distinct data flow 
values reaching each procedure is often very small. This is
especially true for heap-based analyses that use bounded abstractions, due to the
locality of references in recursive paths. This claim is validated is Section~\ref{sec-results},
in which we present the results of a points-to analysis. 

\subsection*{Algorithm}

\renewcommand*\Call[2]{\textproc{#1}(#2)}
\begin{figure}[t]
\begin{algorithmic}[1]
\State \textbf{global} $contexts$, $transitions$, $worklist$
\Procedure{initContext}{$X$}
\State $\Call{add}{contexts, X}$
\State Set $\Call{exitValue}{X} \gets \top$
\State Let $m \gets \Call{method}{X}$
\ForAll{nodes $n$ in the body of $m$}
\State $\Call{add}{worklist, \langle X, n \rangle}$
\State Set $\Call{in}{X,n} \gets \top$ and $\Call{out}{X,n} \gets \top$
\EndFor
\State Set $\Call{in}{X,\Call{entryNode}{m}} \gets \Call{entryValue}{X}$
\EndProcedure
\Procedure{doAnalysis}{}
\State $\Call{initContext}{\langle \texttt{main}, BI \rangle}$
\While{$worklist$ is not empty}
\State Let $\langle X, n \rangle \gets \Call{removeNext}{worklist}$
\If{n is not the entry node}
\State Set $\Call{in}{X,n} \gets \top$
\ForAll{predecessors $p$ of $n$}
\State Set $\Call{in}{X,n} \gets \Call{in}{X,n} \sqcap \Call{out}{X,p}$
\EndFor
\EndIf
\State Let $a \gets \Call{in}{X,n}$
\If{$n$ contains a method call}
\State Let $m \gets \Call{targetMethod}{n}$ 
\State Let $x \gets \Call{callEntryFlowFunction}{X, m, n, a}$
\State Let $X' \gets \langle m, x \rangle$ \Comment $x$ is the entry value at $m$
\State Add an edge $\langle X, n \rangle \rightarrow X'$ to $transitions$
\If{$X' \in contexts$}
\State Let $y \gets \Call{exitValue}{X'}$
\State Let $b_1 \gets \Call{callExitFlowFunction}{X, m, n, y}$
\State Let $b_2 \gets \Call{callLocalFlowFunction}{X, n, a}$
\State Set $\Call{out}{X, n} \gets b_1 \sqcap b_2$ 
\Else
\State $\Call{initContext}{X'}$
\EndIf
\Else
\State Set $\Call{out}{X,n}~\gets~\Call{normalFlowFunction}{X, n, a}$
\EndIf
\If{$\Call{out}{X,n}$ has changed}
\ForAll{successors $s$ of $n$}
\State $\Call{add}{worklist, \langle X, s \rangle}$
\EndFor
\EndIf
\If{$n$ is the exit node}
\State Set $\Call{exitValue}{X} \gets \Call{out}{X,n}$
\ForAll{edges $\langle X', c \rangle \rightarrow X$ in $transitions$}
\State $\Call{add}{worklist, \langle X', c \rangle}$
\EndFor
\EndIf
\EndWhile
\EndProcedure	
\end{algorithmic}
\caption{Algorithm for performing inter-procedural analysis using value contexts.}
\label{fig-algorithm}
\end{figure}

Figure~\ref{fig-algorithm} provides the  overall algorithm. 
Line 1 declares three globals: a set of contexts that have been created,
a transition table mapping a context and call site of a caller method
to a target context at the called method and a work-list of context-parametrized
control-flow graph nodes whose flow function has to be processed.

\usetikzlibrary{backgrounds}

\tikzstyle{cfg} = [name=#1, rectangle, draw, 
	minimum width=50, minimum height=15, text centered]
\begin{figure*}[!t]
\begin{tabular}{@{}c|@{}c}
\begin{tabular}{@{}c}
\begin{tikzpicture}[node distance=0.7 and -0.5]
\node[cfg=m1] {\tt main()};
\node[cfg=m2] [below=of m1, label=left:$n_1$] {\tt p = 5};
\node[cfg=m3] [below=of m2, label=left:$c_1$] {\tt q = f(p, -3)};
\node[cfg=m4] [below=of m3, label=left:$c_4$] {\tt r = g(-q)};
\node[cfg=m5] [below=of m4, label=left:$n_6$] {\tt exit};
\draw[->] (m1) to coordinate (m1-m2) (m2);
\draw[->] (m2) to coordinate (m2-m3) (m3);
\draw[->] (m3) to coordinate (m3-m4) (m4);
\draw[->] (m4) to coordinate (m4-m5) (m5);
\node[cfg=f1] [node distance=2.1, right=of m1] {\tt f(a, b)};
\node[cfg=f2] [below=of f1, label=left:$n_2$] {\tt if (...)};
\node[cfg=f3] [below left=of f2, label=left:$n_3$] {\tt c = a * b};
\node[cfg=f4] [below right=of f2, label=left:$c_2$] {\tt c = g(10)};
\node[cfg=f5] [below left=of f4, label=left:$n_4$]       {};
\node[cfg=f6] [below=of f5, label=left:$n_5$]       {\tt return c};
\draw[->] (f1) to coordinate (f1-f2) (f2);
\draw[->] (f2) to coordinate (f2-f3) (f3);
\draw[->] (f2) to coordinate (f2-f4) (f4);
\draw[->] (f3) to coordinate (f3-f5) (f5);
\draw[->] (f4) to coordinate (f4-f5) (f5);
\draw[->] (f5) to coordinate (f5-f6) (f6);
\node[cfg=g1] [node distance=2.1, right=of f1] {\tt g(u)};
\node[cfg=g2] [below=of g1, label=left:$c_3$] {\tt v = f(-u, u)};
\node[cfg=g3] [below=of g2, label=left:$n_6$] {\tt return v};
\draw[->] (g1) to coordinate (g1-g2) (g2);
\draw[->] (g2) to coordinate (g2-g3) (g3);
\begin{scope}[node distance=0.1, text width=1cm, font=\scriptsize]
\node [node distance=0.65, left =of m1-m2] {$\langle X_0, \top \rangle$};
\node [node distance=0.65, left =of m2-m3] {$\langle X_0, p^+ \rangle$};
\node [node distance=0.65, left =of m3-m4] {$\langle X_0, p^+q^- \rangle$};
\node [node distance=0.65, left =of m4-m5] {$\langle X_0, p^+q^-r^- \rangle$};
\node [right=of f1-f2] {$\langle X_1, a^+b^- \rangle$ $\langle X_3, a^-b^+ \rangle$};
\node [node distance=0.8, left =of f2-f3] {$\langle X_1, a^+b^- \rangle$ $\langle X_3, a^-b^+ \rangle$};
\node [right=of f2-f4] {$\langle X_1, a^+b^- \rangle$ $\langle X_3, a^-b^+ \rangle$};
\node [node distance=0.8, left =of f3-f5] {$\langle X_1, a^+b^-c^- \rangle$ $\langle X_3, a^-b^+c^- \rangle$};
\node [right=of f4-f5] {$\langle X_1, a^+b^-c^- \rangle$ $\langle X_3, a^-b^+c^- \rangle$};
\node [right=of f5-f6]  {$\langle X_1, a^+b^-c^- \rangle$ $\langle X_3, a^-b^+c^- \rangle$};
\node [right=of g1-g2] {$\langle X_2, u^+ \rangle$};
\node [right=of g2-g3] {$\langle X_2, u^+v^- \rangle$};
\end{scope}
\end{tikzpicture}
\vspace*{.25cm}
\hspace*{.1cm}
\\
(a) \ \ 
\parbox[t]{100mm}{Control flow graphs annotated with context-sensitive data flow values}
\end{tabular}
&
\begin{tabular}{c}
\begin{tabular}{c}
\begin{tikzpicture}[node distance=0.3 and 0.3]
\node[name=top] {$\top$};
\node[below left=of top, name=minus] {$-$};
\node[below=of top, name=zero] {$0$};
\node[below right=of top, name=plus] {$+$};
\node[below=of zero, name=bot] {$\bot$};
\draw (top) to (minus);
\draw (top) to (zero);
\draw (top) to (plus);
\draw (minus) to (bot);
\draw (zero) to (bot);
\draw (plus) to (bot);
\end{tikzpicture}
\\
(b) \ \ Lattice for a single variable\rule[-.7em]{0em}{.9em}
\end{tabular}
\\ \hline\rule{0em}{3.75em}
\hspace*{.1cm}
\begin{tabular}{|c|c|c|c|}
\hline
Context & Proc. & Entry & Exit \\
\hline
\rule{0em}{1em}%
$X_0$ & \texttt{main} & $\top$ & $p^+q^-r^-$ \\
$X_1$ & \texttt{f} & $a^+b^-$ & $a^+b^-c^-$ \\
$X_2$ & \texttt{g} & $u^+$ & $u^+v^-$ \\
$X_3$ & \texttt{f} & $a^-b^+$ & $a^-b^+c^-$ \\
\hline
\end{tabular}
\\
(c) \ \ Value contexts for the program\rule[-.6em]{0em}{2em}
\\ \hline
\begin{tabular}{c}
\begin{tikzpicture}[node distance=0.5]
\node[name=X0] {$X_0$};
\node[name=X1, right=of X0] {$X_1$};
\node[name=X2, right=of X1] {$X_2$};
\node[name=X3, right=of X2] {$X_3$};
\draw[->] [bend left] (X0) to node [above] {$c_1$} (X1);
\draw[->] [bend left] (X1) to node [above] {$c_2$} (X2);
\draw[->] [bend left]  (X2) to node [above] {$c_3$}  (X3);
\draw[->] [bend left] (X3) to node [below] {$c_2$}  (X2);
\draw[->] [bend right] (X0) to node [below] {$c_4$}  (X2);
\end{tikzpicture}
\\
(d) \ \ Context transition diagram 
\end{tabular}
\end{tabular}
\end{tabular}
\caption{A motivating example of a non-distributive sign-analysis performed on a program with mutually recursive procedures.}
\label{fig-cfg}
\end{figure*}
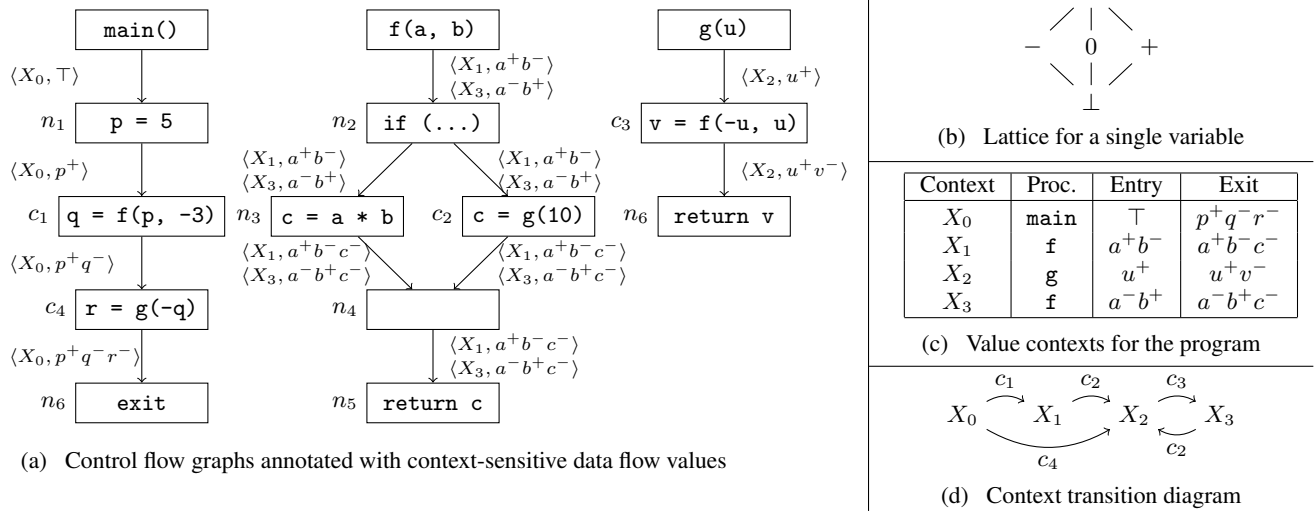

The procedure \textsc{initContext} (lines 2-11) initializes a new context
with a given method and entry value. The exit value is initialized 
to the $\top$ element. IN/OUT values at all nodes in the
method body are also initialized to $\top$, with the exception of the 
method's entry node, whose IN value is initialized to the context's entry value. 
All nodes of this context are added to the work-list.

The \textsc{doAnalysis} procedure (lines 12-51) first creates 
a value context for the \texttt{main} method with some 
boundary information (BI). Then, data flow analysis is performed using
the traditional work-list method, but distinguishing between nodes of different
contexts.

A node is removed from the work-list and its IN value is set to the meet of
the OUT values of its predecessors (lines 16-21). For nodes without a method call,
the OUT value is computed using the normal flow function (line 37).
For call nodes, parameter passing is handled by a call-entry flow function that takes as input
the IN value at the node, and the result of which is used as the entry value at the 
callee context (lines 24-26). The transition
from caller context and call-site to callee context is also recorded (line 27).
If a context with the target method and computed entry value has not been
previously created, then it is initialized now (line 34). Otherwise, the exit value
of the target context is used as the input to a call-exit flow function, to handle returned values. 
A separate call-local flow function takes as input the IN value at the call node, and propagates information
about local variables. 
The results of these two functions are merged into the OUT value of the call node (lines 29-32). 

Once a node is processed, its successors are added to 
the work-list if its OUT value has changed in this iteration (lines 39-43). 
If the node is the exit of its procedure 
(lines 44-49), then the exit value of its context is set and all its callers are 
re-added to the work-list.

The termination of the algorithm follows from the monotonicity of flow functions
and the finiteness of the lattice (which bounds the descending chain as well
as the number of value contexts).

The algorithm can easily be extended to handle multiple entry/exit points per procedure
as well as virtual method calls by merging data flow values across these
multiple paths. It can also be easily adapted for backward data flow analyses.

\subsection*{Example}

Consider the program in Figure~\ref{fig-cfg}~(a), for which we wish to perform
a simplified \emph{sign analysis}, to determine whether a scalar local variable
is negative, positive or zero. The call from \texttt{main} to \texttt{f} at $c_1$ 
will only return when the mutual recursion of \texttt{f} and \texttt{g} terminates,
which happens along the program path $n_2n_3n_4n_5$. Notice that the arguments
to \texttt{f} at call-site $c_3$ are always of opposite signs, causing the value of 
variable $c$ to be negative after every execution of $n_3$ in this context. 
Thus, \texttt{f} and hence \texttt{g} always returns a negative value. 

To compute this result using the algorithm described above, we use data flow
values that are elements of
the lattice in Figure~\ref{fig-cfg}~(b),
where $\top$ indicates an uninitialized variable and $\bot$ is the conservative 
assumption. We use superscripts to map variables to a sign or $\bot$,
and omit uninitialized variables. 

At the start of the program no variables are initialized and hence the
analysis starts with the initial value context 
\text{$X_0 = \langle \texttt{main}, \top \rangle$}. For
work-list removal, we will use lexicographical ordering of contexts (newer first)
before nodes (reverse post-order).

The flow function of $\langle X_0, n_1 \rangle$ is processed first, which
makes $p$ positive (written as $p^+$). The next node picked from the work-list is $c_1$, whose
call-entry flow function passes one positive and one negative
argument to parameters $a$ and $b$ of procedure \texttt{f} respectively. 
Thus, a new value context \text{$X_1 = \langle \texttt{f}, a^+b^- \rangle$} is created
and the transition $\langle X_0, c_1 \rangle \rightarrow X_1$ is recorded. 

Analysis proceeds by processing $\langle X_1, n_2 \rangle$ and then 
$\langle X_1, c_2 \rangle$, which creates a new value context 
$X_2 = \langle \texttt{g}, u^+ \rangle$ due to the positive
argument. The transition $(X_1, c_2) \rightarrow X_2$ is recorded.
When $\langle X_2, c_3 \rangle$ is processed, the arguments to \texttt{f} are
found to be negative and positive respectively, creating
a new value context $X_3 = \langle \texttt{f}, a^-b^+ \rangle$ and a 
transition $(X_2, c_3) \rightarrow X_3$.

The work-list now picks nodes of context $X_3$, and when $\langle X_3, c_2 \rangle$ is
processed, the entry value at \texttt{g} is $u^+$,
for which a value context already exists -- namely $X_2$. The transition
\text{$\langle X_3, c_2 \rangle \rightarrow X_2$} is recorded.
The exit value of $X_2$ is at the moment $\top$ because its exit node has not 
been processed. Hence, the call-exit flow function determines the
returned value to be uninitialzed and the OUT of $\langle X_3, c_2 \rangle$ gets the
value $a^-b^+$. The next 
node to be processed is $\langle X_3, n_3 \rangle$, whose flow function computes the sign
of $c$ to be negative as it is the product of a negative and positive value.
The IN value at $\langle X_3, n_4 \rangle$ is ($a^-b^+c^- \sqcap a^-b^+) = a^-b^+c^-$.
Thus, the sign of the returned variable $c$ is found to be negative. As $n_4$ 
is the exit node of procedure \texttt{f}, the callers of $X_3$ are looked up 
in the transition table and added to the work-list.

\begin{figure*}[!t]
	\begin{center}
	\includegraphics[scale=0.6]{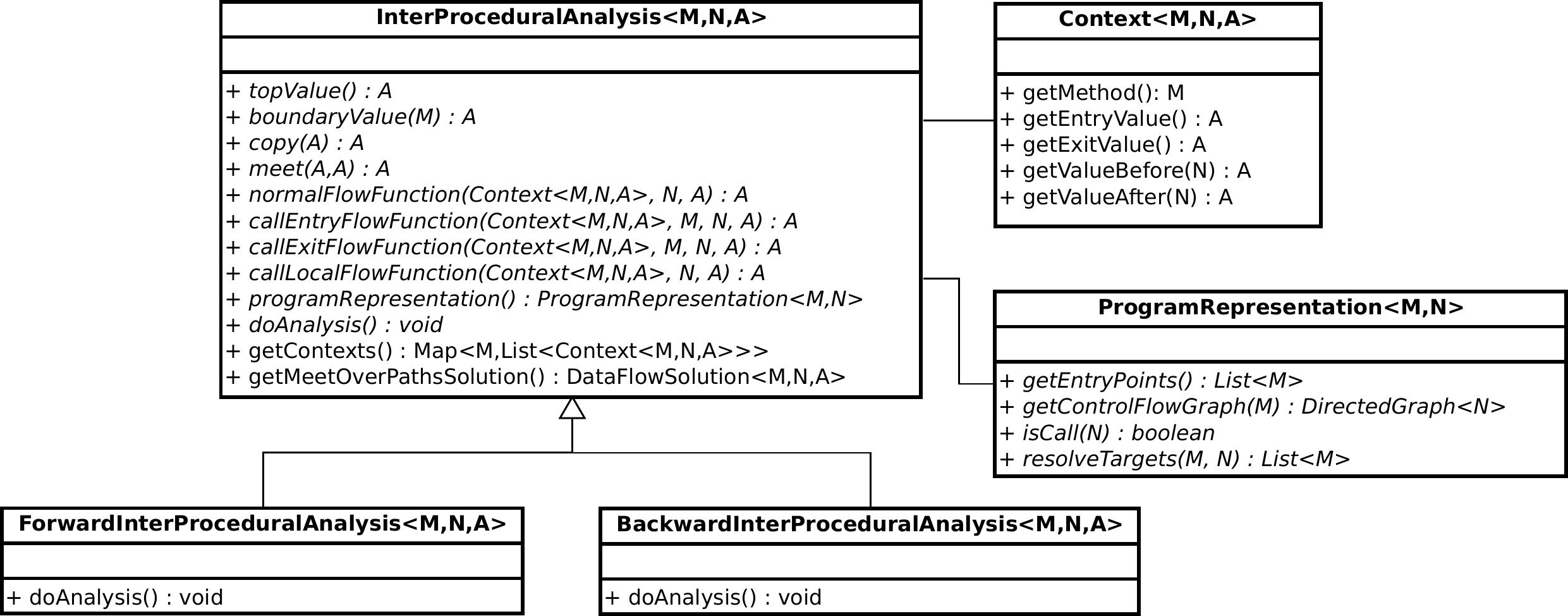}
	\end{center}
	\caption{The class diagram of our generic interprocedural analysis framework.}
	\label{fig-ipa-classes}
\end{figure*}

The only caller $\langle X_2, c_3 \rangle$ is now re-processed, this time
resulting in a hit for an existing target context $X_3$. The exit value of $X_3$
being $a^-b^+c^-$, the returned variable $v$ gets a negative sign, 
which propagates to the exit node $n_6$. The callers of 
$X_2$, namely $\langle X_1, c_2 \rangle$ and $\langle X_3, c_2 \rangle$, are
re-added to the work-list.

$\langle X_3, c_2 \rangle$ is processed next, and this time the correct exit
value of target context $X_2$, which is $u^+v^-$, is used and the OUT
of $\langle X_3, c_2 \rangle$ is set to $a^-b^+c^-$. 
When its successor $\langle X_3, n_4 \rangle$
is subsequently processed, the OUT value does not change and hence no more nodes
of $X_3$ are added to the work-list. Analysis continues with nodes of $X_1$ on
the work-list, such as $\langle X_1, c_2 \rangle$ and $\langle X_1, n_3 \rangle$. 
The sign of $c$ is determined to be negative and this propagates to the 
end of the procedure. When exit node $\langle X_1, n_5 \rangle$ is processed, the
caller of $X_1$, namely $\langle X_0, c_1 \rangle$, is re-added to the work-list.
Now, when this node is processed, $q$ is found to be negative.

Value-based contexts are not only useful in 
terminating the analysis of recursive procedures, as shown above, but also
as a simple \emph{cache} table for distinct call sites. For example,
when $\langle X_0, c_4 \rangle$ is processed, the positive argument results in a hit
for $X_2$, and thus its exit value is simply re-used to determine that $r$ is negative.
Figure~\ref{fig-cfg}~(b) lists the value contexts for the program and 
Figure~\ref{fig-cfg}~(c) shows the transitions between contexts at call-sites.

A context-insensitive analysis would have merged signs of $a$ and $b$ across
all calls to \texttt{f} and would have resulted in a $\bot$ value for the
signs of $c$, $v$, $q$ and $r$. Our context-sensitive method ensures a precise data flow solution
even in the presence of recursion.

Notice that the flow function for $n_3$ is non-distributive since
$f_{n_3}(a^+b^-) \sqcap f_{n_3}(a^-b^+) = a^+b^-c^- \sqcap a^-b^+c^- = a^{\bot}b^{\bot}c^-$
but $f_{n_3}(a^+b^- \sqcap a^-b^+) = f_{n_3}(a^{\bot}b^{\bot}) = a^{\bot}b^{\bot}c^{\bot}$.
Hence this problem does not fit in the IFDS/IDE framework, but such flow functions
do not pose a problem to our algorithm.

\section{Implementation Framework}
\label{sec-framework}

The implementation framework consists of a handful of core
classes 
as shown in Figure~\ref{fig-ipa-classes}. The use of generic
types makes the framework agnostic to any particular toolkit or IR. The
classes are parameterized by three types: \texttt{M} represents the type
of a method, \texttt{N} represents a node in the control flow graph
and \texttt{A} is the type
of data flow value used by the client analysis.
The framework can be naturally instantiated for Soot using the type 
parameters \texttt{SootMethod} and \texttt{Unit} for \texttt{M} and \texttt{N}
respectively.

Users would extend \texttt{ForwardInterProceduralAnalysis} or
\texttt{BackwardInterProceduralAnalysis}, which are subclasses of an abstract
class \texttt{InterProceduralAnalysis}. The abstract methods  
\texttt{topValue}, 
\texttt{boundaryValue},
\texttt{copy} and \texttt{meet} provide a 
hook for client analyses to express initial lattice values and basic
operations on them. The major functionality of the client analysis would be
present in the \texttt{*FlowFunction} methods, whose roles were explained in
Section~\ref{sec-background}. Additionally, clients are expected to provide a 
\texttt{ProgramRepresentation} object, which specifies program entry points 
(for which boundary values are to be defined) and resolves virtual calls. 
Our framework ships with default program
representations for Soot's Jimple IR. The launch point of
the analysis is the \texttt{doAnalysis} method, which is
implemented as per the algorithm from Figure~\ref{fig-algorithm} in the
directional sub-classes.

The \texttt{Context} class encapsulates information about a
value context. Every context is associated with a method, an entry value
and an exit value, each of which can be retrieved using the corresponding
\emph{getter} methods. The \texttt{getValueBefore} and \texttt{getValueAfter}
methods return data flow values for a context just before and after a node
respectively. This is the recommended way for accessing the results of the
analysis in a context-sensitive manner. A mapping of methods to a list of
all its contexts is available through the \texttt{getContexts}
method of the \texttt{InterProceduralAnalysis} class. Alternatively, 
\texttt{getMeetOverValidPathsSolution} can be used to obtain a solution that is
computed by merging data flow results across all contexts of each method. The
\texttt{DataFlowSolution} class (not shown in the figure) simply provides 
\texttt{getValueBefore} and \texttt{getValueAfter} methods to access the resulting
solution.

\section{The Role of Call Graphs}
\label{sec-results}

We initially developed this framework in order to implement heap reference 
analysis~\cite{heap-reference-analysis} using Soot, because it could not
be encoded as an IFDS/IDE problem.
However, even with our general framework, performing whole-program analysis
turned out to be infeasible due to a large number of interprocedural paths
arising from conservative assumptions for targets of virtual calls.

The \textsc{spark} engine~\cite{spark} in Soot uses a flow and context 
insensitive pointer analysis on the whole program to build the call graph,
thus making conservative assumptions for the targets of virtual calls in 
methods that are commonly used such as those in the Java library.  
For example, it is not uncommon to find call sites in library methods with 5 or 
more targets, most of which will not be traversed in a given context. Some
call sites can even be found with more than 250 targets! This is common with calls
to virtual methods defined in \texttt{java.lang.Object}, such as \texttt{hashCode()}
or \texttt{equals()}.  

When performing whole-program data flow analysis, 
the use of an imprecise call graph hampers both efficiency,
due to an exponential blow-up of spurious paths, and precision, due to 
the meet over paths that are actually interprocedurally invalid, thereby
diminishing the gains from context-sensitivity. 

Soot provides a context-sensitive call graph builder called \textsc{paddle} \cite{paddle},
but this framework can only perform $k$-limited call-site or object-sensitive
analysis, and that too in a flow-insensitive manner. We were unable to use 
\textsc{paddle} with our framework directly because at the moment it not clear to us 
how the $k$-suffix contexts of \textsc{paddle} would map to our value-contexts.

\begin{table*}[t]
\centering
\begin{tabular}{|c|l|r|r|r|r|r|r|r|r|r|}
\hline
\multicolumn{2}{|c|}{\multirow{2}{*}{Benchmark}} & \multirow{2}{*}{Time} & \multicolumn{2}{c|}{Methods ($M$)} & \multicolumn{2}{c|}{Contexts ($X$)} & \multicolumn{2}{c|}{$X/M$} & \multicolumn{2}{c|}{Clean}\\
\cline{4-11}
\multicolumn{2}{|c|}{} & & Total & App. & Total & App. & Total & App. & Total & App.\\
\hline
\multirow{5}{*}{SPEC JVM98} 
& \texttt{compress} & 1.15s &  367 & 54 & 1,550 & 70 & 4.22 & 1.30 & 50 & 47 \\
& \texttt{jess} & 140.8s &  690 & 328 & 17,280 & 9,397 & 25.04 & 28.65 & 34 & 30 \\
& \texttt{db} & 2.19s &  420 & 56 & 2,456 & 159 & 5.85 & 2.84 & 62 & 46 \\
& \texttt{mpegaudio} & 4.51s &  565 & 245 & 2,397 & 705 & 4.24 & 2.88 & 50 & 47 \\
& \texttt{jack} & 89.33s &  721 & 288 & 7,534 & 2,548 & 10.45 & 8.85 & 273 & 270 \\
\hline
\multirow{2}{*}{DaCapo 2006}
& \texttt{antlr} & 697.4s &  1,406 & 798 & 30,043 & 21,599 & 21.37 & 27.07 & 769 & 727 \\
& \texttt{chart} & 242.3s &  1,799 & 598 & 16,880 & 4,326 & 9.38 & 7.23 & 458 & 423 \\
\hline
\end{tabular}
\caption{Results of points-to analysis using our framework. ``App.'' refers to data for application classes only.}
\label{tab-methods}
\end{table*}

\subsection*{Call Graph Construction using Points-To Analysis}

We have implemented a flow and context-sensitive points-to 
analysis using our interprocedural framework to build a call graph 
on-the-fly. This analysis is both a demonstration of the use of our framework as
well as a proposed solution for better call graphs intended for use by other
interprocedural analyses.

The data flow value used in our analysis is a points-to graph in which nodes are allocation
sites of objects. We maintain two types of edges: $x \rightarrow m$ 
indicates that the root variable $x$ may point to objects allocated at site
$m$, and $m.f \rightarrow n$ indicates that objects allocated at site $m$
may reference objects allocated at site $n$ along the field $f$. 
Flow functions add or remove edges when processing assignment statements
involving reference variables. Nodes that become unreachable from root
variables are removed. Type consistency is maintained by propagating only
valid casts.

The points-to graphs at each statement only maintain objects reachable
from variables that are local to the method containing the statement. At
call statements, we simulate assignment of arguments to locals of the 
called method, as well as the assignment of returned values to a local
of the caller method. For static fields (and objects reachable from them)
we maintain a global flow-insensitive points-to graph. For statements involving
static loads/stores we operate on a temporary union of local and global
graphs. The call graph is constructed on-the-fly by resolving virtual method
targets using type information of receiver objects.

Points-to information cannot be precise for objects returned
by native methods, and for objects shared between multiple threads (as our
analysis is flow-sensitive). Thus, we introduce the concept of a 
\emph{summary node}, which represents statically unpredictable points-to
information and is denoted by the symbol $\bot$. For soundness, we must
conservatively propagate this effect to variables and fields that involve
assignments to summary nodes. 
The rules
for summarization along different types of assignment statements are as follows:

\begin{center}
\begin{tabular}{|c|l|}
\hline
Statement & Rule used in the flow function \\
\hline
$x = y$   & If $y \rightarrow \bot$, then set $x \rightarrow \bot$ \\
$x.f = y$ & If $y \rightarrow \bot$, then $\forall o : x \rightarrow o$,
                 set $o.f \rightarrow \bot$ \\
$x = y.f$ & If $y \rightarrow \bot$ or $\exists o : y \rightarrow o$ and $o.f \rightarrow \bot$,
				 \\ & then set $x \rightarrow \bot$ \\
$x = p(a_1, a_2, ...)$ & If $p$ is unknown, then set $x \rightarrow \bot$, 
                 and \\ & $\forall o : a_i \rightarrow o$,
                     $\forall f \in fields(o)$ set $o.f \rightarrow \bot$ \\
\hline
\end{tabular}
\end{center}

The last rule is drastically conservative; for
soundness we must assume that a call to an unknown procedure may modify
the fields of arguments in any manner, and return any object. An important
discussion would be on what constitutes an \emph{unknown} procedure.
Native methods primarily fall into this category.
In addition, if $p$ is a virtual method invoked on a reference variable
$y$ and if $y \rightarrow \bot$, then we cannot determine precisely what the target
for $p$ will be. Hence, we consider this call site as a \emph{default} site,
and do not enter the procedure, assuming worst-case behaviour for its arguments 
and returned values.
A client analysis using the resulting call graph with our framework
can choose to do one of two things when
encountering a \emph{default} call site: 
(1) assume worst case behaviour for its arguments (eg. in liveness analysis, 
assume that all arguments and objects reachable from them are live)
and carry on to the next statement, or 
(2) fall-back onto
Soot's default call graph and follow the targets it gives. 

A related approach partitions a call graph into calls from 
application classes and library classes~\cite{application-only-call-graph}.
Our call graph is partitioned into call sites that we can precisely resolve 
to one or more valid targets, and those that cannot due 
to statically unpredictable factors.

\subsection*{Experimental Results}

Table~\ref{tab-methods} lists the results of points-to analysis performed on 
seven benchmarks. 
The experiments were carried out on an
Intel Core i7-960 with 19.6 GB of RAM running Ubuntu 12.04 (64-bit) and JDK version 1.6.0\_27. 
Our single-threaded analysis used only one core.

The first two columns contain the names of the benchmarks;
five of which are the single-threaded programs from the SPEC JVM98 suite \cite{specjvm98}, 
while the last two are from the DaCapo suite \cite{dacapo} version 2006-10-MR2.
The third column contains the time required to perform our analysis, which
ranged from a few seconds to a few minutes.
The fourth and fifth columns contain the number of methods analyzed
(total and application methods respectively).
The next two columns contain the number of value-contexts created, with the
average number of contexts per method in the subsequent two columns. It can be seen
that the number of distinct data flow values reaching 
a method is not very large in practice. 
As our analysis ignores paths with method invocations on null pointers, it was
inappropriate for other benchmarks in the DaCapo suite when using 
stub classes to simulate the suite's reflective boot process.

The use of \emph{default} sites in our call graph has two consequences: (1)~the 
total number of analyzed methods may be less than the total number of 
reachable methods and (2)~methods reachable from \emph{default} call sites
(computed using \textsc{spark}'s call graph) cannot be soundly optimized by a client analysis
that jumps over these sites. The last column lists the number of 
\emph{clean} methods which are not reachable from \emph{default} sites and hence can
be soundly optimized. In all but two cases, the majority of application methods are clean.

\begin{table*}[t]
\centering
\begin{tabular}{|c|r|r|r|r|r|r|r|r|r|r|r|}
\hline
\multicolumn{2}{|c|}{Depth $k=$} & \textbf{1} & \textbf{2} & \textbf{3} & \textbf{4} & \textbf{5} & \textbf{6} & \textbf{7} & \textbf{8} & \textbf{9} & \textbf{10} \\
\hline
\multirow{3}{*}{\texttt{compress}} & 
	  \textbf{\textsc{fcpa}} & 2 & 5 & 7 & 20 & 55 & 263 & 614 & 2,225 & 21,138 & 202,071 \\ 
	& \textbf{\textsc{spark}} & 2 & 5 & 9 & 22 & 57 & 273 & 1,237 & 23,426 & 545,836 & 12,052,089 \\ 
	& \textbf{$\Delta$\%} & \textbf{0} & \textbf{0} & \textbf{22.2} & \textbf{9.09} & \textbf{3.51} & \textbf{3.66} & \textbf{50.36} & \textbf{90.50} & \textbf{96.13} & \textbf{98.32} \\ 
\hline
\multirow{3}{*}{\texttt{jess}} & 
	  \textbf{\textsc{fcpa}} & 2 & 5 & 7 & 30 & 127 & 470 & 4,932 & 75,112 & 970,044 & 15,052,927 \\ 
	& \textbf{\textsc{spark}} & 2 & 5 & 9 & 32 & 149 & 924 & 24,224 & 367,690 & 8,591,000 & 196,801,775 \\ 
	& \textbf{$\Delta$\%} & \textbf{0} & \textbf{0} & \textbf{22.2} & \textbf{6.25} & \textbf{14.77} & \textbf{49.13} & \textbf{79.64} & \textbf{79.57} & \textbf{88.71} & \textbf{92.35} \\ 
\hline
\multirow{3}{*}{\texttt{db}} & 
	  \textbf{\textsc{fcpa}} & 2 & 5 & 11 & 46 & 258 & 1,791 & 21,426 & 215,465 & 2,687,625 & 42,842,761 \\ 
	& \textbf{\textsc{spark}} & 2 & 5 & 13 & 48 & 443 & 4,726 & 71,907 & 860,851 & 13,231,026 & 245,964,733 \\ 
	& \textbf{$\Delta$\%} & \textbf{0} & \textbf{0} & \textbf{15.4} & \textbf{4.17} & \textbf{41.76} & \textbf{62.10} & \textbf{70.20} & \textbf{74.97} & \textbf{79.69} & \textbf{82.58} \\ 
\hline
\multirow{3}{*}{\texttt{mpegaudio}} & 
	  \textbf{\textsc{fcpa}} & 2 & 14 & 42 & 113 & 804 & 11,286 & 129,807 & 1,772,945 & 27,959,747 & 496,420,128 \\ 
	& \textbf{\textsc{spark}} & 2 & 16 & 46 & 118 & 834 & 15,844 & 250,096 & 4,453,608 & 87,096,135 & 1,811,902,298 \\ 
	& \textbf{$\Delta$\%} & \textbf{0} & \textbf{12} & \textbf{8.7} & \textbf{4.24} & \textbf{3.60} & \textbf{28.77} & \textbf{48.10} & \textbf{60.19} & \textbf{67.90} & \textbf{72.60} \\ 
\hline
\multirow{3}{*}{\texttt{jack}} & 
	  \textbf{\textsc{fcpa}} & 2 & 18 & 106 & 1,560 & 22,652 & 235,948 & 2,897,687 & 45,480,593 & 835,791,756 & 17,285,586,592 \\ 
	& \textbf{\textsc{spark}} & 2 & 18 & 106 & 1,577 & 27,201 & 356,867 & 5,583,858 & 104,211,833 & 2,136,873,586 & 46,356,206,503  \\ 
	& \textbf{$\Delta$\%} & \textbf{0} & \textbf{0} & \textbf{0} & \textbf{1.08} & \textbf{16.72} & \textbf{33.88} & \textbf{48.11} & \textbf{56.36} & \textbf{60.89} & \textbf{62.71} \\ 
\hline
\multirow{3}{*}{\texttt{antlr}} & 
	  \textbf{\textsc{fcpa}} & 6 & 24 & 202 & 560 & 1,651 & 4,669 & 18,953 & 110,228 & 975,090 & 11,935,918 \\ 
	& \textbf{\textsc{spark}} & 6 & 24 & 206 & 569 & 1,669 & 9,337 & 107,012 & 1,669,247 & 27,670,645 & 468,973,725 \\ 
	& \textbf{$\Delta$\%} & \textbf{0} & \textbf{0} & \textbf{1.9} & \textbf{1.58} & \textbf{1.08} & \textbf{49.99} & \textbf{82.29} & \textbf{93.40} & \textbf{96.48} & \textbf{97.45} \\ 
\hline
\multirow{3}{*}{\texttt{chart}} & 
	  \textbf{\textsc{fcpa}} & 6 & 24 & 217 & 696 & 2,109 & 9,778 & 45,010 & 517,682 & 7,796,424 & 164,476,462 \\ 
	& \textbf{\textsc{spark}} & 6 & 24 & 219 & 714 & 2,199 & 20,171 & 306,396 & 7,676,266 & 192,839,216 & 4,996,310,985 \\ 
	& \textbf{$\Delta$\%} & \textbf{0} & \textbf{0} & \textbf{0.9} & \textbf{2.52} & \textbf{4.09} & \textbf{51.52} & \textbf{85.31} & \textbf{93.26} & \textbf{95.96} & \textbf{96.71} \\ 
\hline
\end{tabular}
\caption{Number of $k$-length call graph paths for various benchmarks using \textsc{spark} and \textsc{fcpa} (Flow and Context-sensitive Pointer Analysis).}
\label{tab-paths}
\end{table*}

In order to highlight the benefits of using the resulting call graph, just
listing the number of edges or call-sites alone is not appropriate,
as our call graph is context-sensitive. 
We have thus computed the number distinct paths in the call graph,
starting from the entry point, which are listed in Table~\ref{tab-paths}.
As the total number of call graph paths is possibly infinite (due to recursion), 
we have counted paths of a fixed length length $k$, for $1 \le k \le 10$.
For each benchmark, we have counted these paths using call graphs constructed
by our Flow and Context-sensitive Pointer Analysis (\textsc{fcpa}) as well
as \textsc{spark}, and noted the difference as percentage savings ($\Delta\%$)
from using our context-sensitive call graph.
The option \texttt{implicit-entry} was set to \texttt{false} for \textsc{spark}.

The savings can be clearly observed for $k > 5$. For $k = 10$,
\textsc{spark}'s call graph contains more than 96\% spurious paths for three 
of the benchmarks, and 62-92\% for the remaining. The gap only widens for larger
values of $k$ (for which the number of paths was too large to compute in some cases).

Client analyses using our interprocedural framework can be configured to use 
our context-sensitive call graphs which avoid these spurious paths, hence
enabling efficient and precise solutions.

\section{Conclusion and Future Work}
\label{sec-conclusion}

We have presented a framework for performing value-based context-sensitive
inter-procedural analysis in Soot. This framework does not require 
distributivity of flow functions and is thus applicable to a large class of
analyses including those that cannot be encoded as IFDS/IDE problems. Another 
advantage of our method is the context-sensitive nature of the resulting data
flow solution, which can be useful in dynamic optimizations.

In order to deal with the difficulties in whole-program analysis performed over an imprecise
call graph, we constructed call graphs on-the-fly while performing
a flow and context-sensitive points-to analysis. This analysis also demonstrated
a sample use of our framework and showed that it was practical to use data flow values as contexts
because the number of distinct data flow values reaching each method
is often very small.

The interprocedural framework has been released and is available
at \url{https://github.com/rohanpadhye/vasco}. However, our points-to analysis implementation is experimental
and makes heavy use of \texttt{HashMap}s and \texttt{HashSet}s, thus running out of memory for
very large programs. We would like to improve this implementation by using
bit-vectors or maybe even BDDs for compact representation of points-to sets.

The precision of our call graphs is still limited in the presence of
\emph{default} sites, which are prominent due to the liberal
use of \emph{summary} nodes in the points-to graphs. We would like to reduce
the number of \emph{summary} nodes by simulating some commonly used native
methods in the Java library, and also by preparing a summary of initialized
static fields of library classes. 
Our hope is that these extensions would 
enable our call graph to be fully complete, thereby enabling users to 
precisely perform whole-program analysis and optimization for all application
methods in an efficient manner. 
We believe that improving the precision of program analysis actually
helps to improve its efficiency, rather than hamper it.

\acks

We would like to thank the anonymous reviewers for suggesting the inclusion of a 
non-distributive example and the separation of call/return flow functions.
These changes improved the paper.


\bibliographystyle{abbrvnat}


\bibliography{mtp.bib}

\newpage

\end{document}